# *Electron and Hole Injection via Charge Transfer at the Topological-Insulator $Bi_{2-x}Sb_xTe_{3-y}Se_y$/Organic-Molecule Interface*

*Yoichi Tanabe*[*1], *Khuong Kim Huynh*[*2], *Ryo Nouchi*[2,3], *Satoshi Heguri*[2], *Gang Mu*[1], *Jingtao Xu*[2], *Hidekazu Shimotani*[1], *Katsumi Tanigaki*[*1,2]

[1]Department of Physics, Graduate School of Science, Tohoku University, 6-3 Aoba, Aramaki, Aoba-ku, Sendai, 980-8578, Japan, [2]WPI-Advanced Institute of Materials Research, Tohoku University, 2-1-1 Katahira, Aoba-ku, Sendai, 980-8577, Japan, [3]Nanoscience and Nanotechnology Research Center, Osaka Prefecture University, Sakai 599-8570, Japan

ABSTRACT

As a methodology for controlling the carrier transport of topological insulators (TI's), a flexible tuning in carrier number on the surface states (SS's) of three dimensional TI's by surface modifications using organic molecules is described. The principle of the carrier tuning and its type conversion of TI's presented in this research are based on the charge transfer of holes or electrons at the TI/organic molecule interface. By employing 2,3,5,6-tetrafluoro-7,7,8,8-tetracyanoquinodimethane (F4-TCNQ) as an electron acceptor or tetracyanoquinodimethane (TCNQ) as a donor for n- and p- $Bi_{2-x}Sb_xTe_{3-y}Se_y$ (BSTS) single crystals, successful carrier conversion from n to p and its reverse mode is demonstrated depending on the electron affinities of the molecules. The present method provides a nondestructive and efficient method for local tuning in carrier density of TI's, and is useful for future applications.

**INTRODUCTION**

A topological insulator (TI) is a material that behaves as an insulator as a bulk state (BS), while permitting metallicity on its Dirac cone (DC) surface state (SS).[1,2] In the bulk of a TI, the electronic band structure resembles an ordinary band insulator, with the Fermi level falling between the conduction and the valence bands. However, on the surface of a topological insulator, the special DC-SS that falls within the bulk energy gap has its spin-momentum locked via a topological order known as a Z2 topological invariant, leading it to be a massless relativistic particle with high metallicity. TI's can generally be obtained using a band inversion with keeping the time reversal symmetry, and were actually observed to occur in a quantum well of mercury telluride cadmium telluride in 2007.[3] TI's were also realized in three dimensional (3D) bulk solids of binary compounds involving elements having large spin-orbit coupling strength, such as bismuth antimony, bismuth selenide, bismuth telluride and antimony telluride,[3,4] and these were experimentally confirmed by angle-resolved photoemission spectroscopy (ARPES).[5] Although the SS's of 3D TIs are expected to be a new type of a two dimensional electron gas (2DEG) system, where an electron spin is locked in its linear momentum, the Fermi levels in many TIs unfortunately fall in either the conduction or the valence bands due to the naturally occurring defects and these must be controlled by external doping.

Since transport measurements on SS's are essential for understanding the fundamental aspects of 3D TIs, methodologies for controlling carrier numbers in both the BS's and the SS's are very important. So far, a fairly small number of reports on electric transport have also been compared to spectroscopic observations. Conventional elemental substitution techniques are reported to change the carrier numbers of SS's from n- to p-type in $Ca_xBi_{2-x}Se_3$, $Bi_{2-x}Sb_xTe_3$, $Bi_{2-x}Sb_xTe_{3-y}Se_y$ (BSTS), and $Bi_{2-x}Sn_xTe_2Se$.[5-12] By employing a field effect transistor (FET) structure, the gate voltage was shown to control ambipolar carrier transport in $Bi_2Se_3$, $Bi_{2-x}Sb_xTe_3$, and $Bi_2Te_3$ thin films.[13-17] Most recently, similar possibilities were shown to control the carrier numbers using absorption of gas species[9, 18, 19], electric double layer (EDL) construction[20] and organic thin film deposition on SS[21]. Considering the flexible control in carrier number of TI's, surface modifications using organic semiconductor molecules as an interface contact on the SS's of TI's are very promising and of importance.

We report here that the carrier type and the number of SS's can successfully be tuned by interface control of BSTS single crystals with a regulated number of inheritance bulk impurities (~ $10^{16}$ cm$^{-3}$ [8-11]) *via* surface modifications by using organic molecules. We will show that the carrier type can be converted from n-type to p-type and *vice-versa* in BSTS single crystals using tetrafluoro-7,7,8,8-tetracyanoquinodimethane (F4-TCNQ) and tetracyanoquinodimethane (TCNQ) with different electron affinities ($E_{ea}$'s). The present studies clearly show that carrier number control as well as type conversion of TIs are possible by employing organic semiconductors having a variety of $E_{ea}$'s or ionization energies (IE's). A topological p-n junction, a new state of matter[22] for realizing novel phenomena of massless Dirac fermions such as Klein tunneling[23] and Veselago lensing[24], could be searched for via the present methods in the near future.

**EXPERIMENTAL SECTION**

Single crystals of n- and p-type BSTS's (n-BSTS : $x = 0.5$, $y = 1.3$, p-BSTS : $x = 1$, $y = 2$) were grown by a slow-cooling method.[8-10] A single crystal was cut into a rectangular shape using a wire cutter and then was mechanically cleaved to obtain a clean surface. The thickness of the single crystal thin films was around 40 μm (n-BSTS) and 65μm (p-BSTS). After cleaving, the single crystal was attached on a substrate using GE varnish and double-sided adhesive for n-BSTS and p-BSTS, respectively, and finally the electrodes were fabricated using gold paste (Tokuriki-kagaku 6430). For making surface modifications, an F4-TCNQ thin film with a thickness of 80 nm was fabricated on n-BSTS at 65 °C using a thermal evaporator under $10^{-6}$ Pa. TCNQ in acetone saturated solution was deposited on p-BSTS using a spin coating method. Electrical resistivities and Hall coefficients were measured using a 4 probe method before and after the fabrications of organic thin films under $B \leq |9\ \text{T}|$ in $2\ \text{K} \leq T \leq 300\ \text{K}$.

**RESULTS AND DISCUSSION**

**1. Electric transport of n- and p-BSTS's before and after surface modifications**

The temperature dependences of longitudinal electrical resistivities ($\rho_{xx}$'s) of prepared n- and p- BSTS's were measured in order to confirm their quality before making their surface modifications by organic molecules as shown in Figs. 1(a) and (b). Actually, the evolution of their resistivities as a function of $T$ showed semiconducting dependences in the intermediate temperature regime. This is consistent with the previous observations on the $T$ dependences of $\rho_{xx}$ for BSTS.[9] The $\rho_{xx}$ values were saturated [10] in the low $T$ regime below 10 K, and this behavior is in sharp contrast to that observed in the intermediate $T$ range. At low $T$'s, the conductivity of normal bulk impurity bands (IBs) is greatly disturbed by the impurity scatterings, while that of the DC-SS is almost insensitive and therefore will give the most predominant contributions to the conductivity at low $T$'s. The observed saturation in $\rho_{xx}$ at low $T$'s is consistent with the major contribution of DC-SS's, but not with the bulk IBs. The experimental $T$ dependences of conductivities ensured that the BSTS samples with both p- and n- characters have sufficient high quality for the purpose of the present experiments.

By employing the BSTS samples described earlier, the changes in the longitudinal resistivities $\rho_{xx}$'s as well as Hall resistivities $\rho_{yx}$'s were studied at low $T$'s as a function of magnetic fields ($B$'s) in detail after the deposition of F4-TCNQ or TCNQ on their surfaces, and the results are displayed in Figs.1(a)-(d).

Figures 1 (c) and (d) show $B$-dependences of Hall resistivities ($\rho_{yx}$) of n- and p-BSTS's with their surface modifications by F4-TCNQ or TCNQ. As described earlier, both prepared n- and p-BSTS's showed electron-like and hole-like $\rho_{yx}$'s together with non-linear responses against $B$ at 2 K, respectively. Upon modification of the top surface of n-BSTS using F4-TCNQ, non-linear electron-like $\rho_{yx}$ against $B$ was still observed, but a clear indication of a sign change in $\rho_{yx}$ was detected around $B$ = 6.5 T at 2 K as seen in Fig.1(c). Similar experiments were made on p-BSTS using TCNQ as a surface modification layer as shown in Figs.1(b) and (d). It is apparent in this experiment that the original non-linear hole-like $\rho_{yx}$ against $B$ was converted to the one showing nonlinear hole-conduction that was clearly evidenced at 2 K. As will be discussed later, these experiments provide unambiguous evidences that the opposite hole or electron carriers can be generated in the SS's in n- or p-BSTS, via the charge transfer from the molecules.

## 2. Analyses of transport properties in a two carrier model

In order to understand the charge transfer at the BSTS/molecule interfaces clearly, we analyzed the $B$-dependences of $\rho_{yx}$ using a two-carrier semiclassical model. The resulting parameters are summarized in Table 1.

We describe the construction of the BSTS/molecule interfaces first. Given the $E_{ea}$ of BSTS to be around 4 - 5 eV by referring to values of $Bi_2Se_3$[25] and $Bi_2Te_3$[26] in the literature, F4-TCNQ ($E_{ea}$

= 5.25)[27] and TCNQ ($E_{ea}$ =3.0 – 3.6 eV)[28] could be regarded as an acceptor and a donor for BSTS, and accordingly we chose these combinations of materials in the present experiments. Scheme 1 displays the electronic states of BSTS to be modified by deposition of F4-TCNQ as well as TCNQ. In the present experimental setups, F4-TCNQ was deposited via vacuum deposition on the top surface of n-BSTS, while TCNQ was cast as an acetone solution on the surface of p-BSTS. Therefore, the bottom side was also modified with TCNQ in addition to the top surface in the latter case, and consequently the charge transfer from TCNQ should be considered not only on the top but also on the bottom SS's of p-BSTS. This is different from the situation of the F4-TCNQ/n-BSTS interface where only the top SS is decorated. As will be discussed in the following paragraphs, additional hole carriers are generated in n-BSTS after the deposition of F4-TCNQ. On the other hand, p-BSTS was successfully converted to the n-type by the treatment of TCNQ. These are consistent with the charge transfer expected at the BSTS/molecule interfaces described on a basis of $E_{ea}$ in this paragraph.

As described in the previous paragraph together with experimental setup, the observed non-linear $\rho_{yx}$ against $B$ could be interpreted in terms of multi-carriers with different carrier numbers as well as with the markedly differentiated high (Dirac-cone SS) and low (impurity band, IB) mobilities. For as-prepared parent n- and p-BSTS's, both systems can be described using a two-carrier model, generally considering the contributions from SS and IB[8]. Employing a semiclassial two-carrier model, both high (1234 $cm^2V^{-1}S^{-1}$) and low (30 $cm^2V^{-1}S^{-1}$) mobilities of n-type carriers were evaluated for n-BSTS. When similar analyses for p-BSTS were made, p-type carriers with high mobility (525 $cm^2V^{-1}S^{-1}$) and n-type carriers with low mobility (60 $cm^2V^{-1}S^{-1}$) were obtained. Keeping the earlier discussions in mind, the n-type carriers with low mobilities evaluated for both n- and p-BSTS's could be ascribed to IBs. These interpretations could be reasonable because such n-type IB will not change before and after the deposition of organic molecules and additional carriers are generated dominantly by thermal excitations. On the other hand, the n- and p-type carriers with high mobilities observed for n- and p-BSTS's can reasonably be attributed to the DC-SS.

After the deposition of F4-TCNQ on the top surface of n-BSTS, n-type carriers with low mobility were converted to the p-type carriers (41 cm$^2$V$^{-1}$S$^{-1}$), while the n-type carriers with high mobility (1126 cm$^2$V$^{-1}$S$^{-1}$) remained unchanged. Considering the relatively large $E_{ea}$ of F4-TCNQ, it is reasonable to consider that holes can be transferred from F4-TCNQ to the top SS of n-BSTS. The n-type SS, therefore, is expected to be converted to the p-type SS. However, the Hall effect experimentally observed showed an n-type behavior, in contrast with this simple thought. Consequently, the n-type carriers evaluated to be of high mobility could alternatively be assigned to the bottom SS of n-BSTS. Since F4-TCNQ was deposited solely on the top surface by vacuum deposition, such assignment would be reasonable. Although the DC-SS was filled with holes by the surface modifications of F4-TCNQ, its mobility will not be high enough when compared to the bottom DC-SS on a substrate, because the chemical potential of the top DC-SS will be far apart from the neutral Dirac point and the mobility of the Dirac electrons is inversely proportional to the square root of the carrier number.

It could be noted that $\rho_{yx}$ becomes negative under extremely high magnetic fields since the n-characteristic IB with low mobility could be responsible for the final value of $\rho_{yx}$ under such high $B$. It should be noted that in the present analyses, however, the sign of $\rho_{yx}$ is still positive at 9 T. This result seems to be given by the effect of the p-type SS converted from the n-type SS via hole transfer from F4-TCNQ, and the n-type IB and the p-type SS in total were consequently analyzed to be a p-type action in the framework of a two carrier model and this is not intrinsic. A more realistic model may be a three carrier model, but we need more accurate data at higher B for making the resulting interpretations valid and more reliable (see the supporting information).

When TCNQ was deposited via a solution method on the surface of p-BSTS, n-type carriers with both high (307 cm$^2$V$^{-1}$S$^{-1}$) and low (34 cm$^2$V$^{-1}$S$^{-1}$) mobilities were found in similar analyses. It could be interpreted that the top and the bottom p-type SS's are successfully converted to the n-type SS, while the IB is not largely influenced. These results are considered to be reasonable when the modifications of both top and bottom sides are taken into account as explained earlier.

## 3. Structures of organic thin films on BSTS single crystals

As discussed earlier, the SS of n- and p-BSTS's were successfully converted to the opposite type after the deposition of organic molecules, and this is consistent with the scenario of the charge transfer effect caused by the difference of their $E_{ea}$'s. Although it is hard to investigate the accurate interface structure between the organic thin films and the BSTS single crystals in the present experiments, the structural information of organic thin films could be helpful for understanding the present results and X-ray diffraction profiles for n-BSTS and p-BSTS are given along the c-axis direction after deposition of F4-TCNQ or TCNQ in Figs. (a) and (d). While the diffraction peaks at (0 0 $l$) ($l = 3n$) indices for BSTS were dominant in the diffraction profiles[8], some additional peaks developed after deposition of F4-TCNQ or TCNQ as shown in Figs. 3 (b), (c), (e), (f). As compared to the diffraction profiles calculated from the structural parameters of F4-TCNQ[29] and TCNQ[30], the ($h$ 0 0) or (0 0 $l$) ($h,l = 2n$) indices could reasonably be assigned to the diffraction from F4-TCNQ or TCNQ films. Therefore, it is most plausible that crystalline organic thin films with typical orientations form on BSTS single crystal surface, and this crystallographic information is consistent with the scenario of charge transfer depending on the $E_{ea}$'s.

**4. General discussion considering other experimental facts**

We analyse the situation of the BSTS/molecule interfaces which we deduced from the present experiments as follows. When F4-TCNQ or TCNQ is used as a surface modification layer on n- or p-BSTS respectively, carriers are transferred at the interface depending on the difference of their $E_{ea}$'s. In the present experiments, the n- or the p-type SS was converted to the opposite type by leaving the persistent n-type IB as the less influenced BS. As actually expected for n-BSTS/F4-TCNQ, a hole-like $\rho_{yx}$ was observed at low $B$ which evolves with a convex shape as $B$ increases due to the influence of electron-like IB with low mobility. On the other hand, for p-BSTS/TCNQ, an electron-like $\rho_{yx}$ was observed at low $B$ which evolves with a concave shape as $B$ increases also under the influence of IB.

It was reported by ARPES that the Dirac point (DP) of charge neutrality in BSTS, having a similar stoichiometry to that of the present sample, is close to the valence band maxima.[11] When the chemical potential resides in the vicinity of DP, the dominant contributions to $\rho_{yx}$ both from n-/p-type SS and from the BS of the electron-like IB via thermal activated carriers from the valence band should be taken into consideration. Our present analyses of the measurements of $\rho_{yx}$ for both as-prepared n- and p-BSTS's clearly demonstrated that n-type carriers with low mobility exist and such evidence is consistent with the band picture previously reported. On the other hand, in the case of F4-TCNQ on n-BSTS, a clear sign change in $\rho_{yx}$ was observed at 2 K around $B = 6.5$ T. Moreover, a hole-like $\rho_{yx}$ was converted to an electron-like $\rho_{yx}$ when TCNQ was used on p-BSTS as shown in Fig.1(d). These experimental data are unambiguous, strong evidence that the SS's in n- and p-BSTS's are now compensated and converted by the p- and n-carriers, respectively, which are additionally transferred from F4-TCNQ or TCNQ, finally leading to a carrier polarity change of the DC-SS's. It is important to comment again that the situation of the TCNQ can be compared to the fact that the inheritant n-type DC-SS was still left with being less perturbed in the opposite bottom SS in the case of F4-TCNQ.

The debate still continues about the possible trivial quantum well (TQW) states in 3D-TI coexisting with the DC[9,31-34]. Although the present results could be understood as the charge transfer at the BSTS/molecule interfaces, which can occur in principle depending on the difference in $E_{ea}$, they can hardly differentiate the contributions between DC and TQW on the SS's solely from the experiments and the interpretations of $\rho_{yx}$. Especially in the case of n-BSTS/F4-TCNQ, as discussed earlier, the present experimental results cannot be understood by a simple description of a chemical potential shift. Other possible interpretations that the hole transfer from F4-TCNQ to the SS of n-BSTS include both DC and TQW states may also be taken into account.

**SUMMARY**

We succeeded to convert the carrier polarity of the SS of BSTS by employing the charge transfer occurring at the BSTS/organic molecule interface. Depending on the difference of $E_{ea}$, n- or p-type carriers are successfully transferred from TCNQ or F4-TCNQ to p-\n-BSTS's, respectively. The present results demonstrated that the charge transfer at TI/organic-molecule interfaces is a very efficient way to control the carrier density of TI's. By employing lithographic techniques or electrostatic lamination of organic thin films on the surface of TIs as a feasible tool for carrier control, we will be able to modify the carrier numbers on DC-SS's without any serious damage to the surfaces of TIs.

A topological p-n junction is currently considered to be a new state of matter[22] which can create novel phenomena via attractive characteristics of massless Dirac fermions, such as Klein tunneling[23] and Veselago lensing.[24] Moreover, a single chiral edge mode can be expected along the boundary of a p-n junction when the time reversal symmetry is broken, which will provide a new platform of spintronics. We could access to these novel electronic states via local deposition of organic molecules on the surface of TIs. By considering a variety of organic molecules with a variety of Ea's and their structural flexibility, we can accurately adjust the chemical potential of TIs in the vicinity of DP. The present method can be a technological platform to realize ultralow-dissipative carrier transport. These could be simple but advantageous to study the frontier areas of TIs associated with magnetoelectric effects[35, 36] in SS transport.

SCHEME

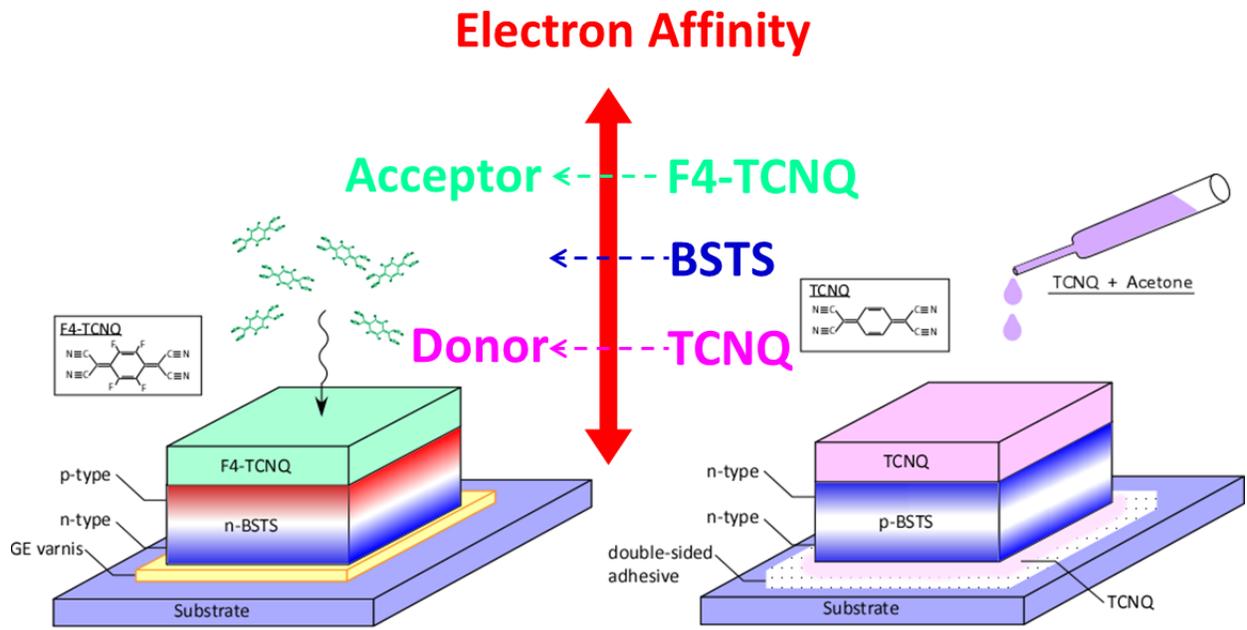

**Scheme1.** A summary of the electronic structure of n- and p- type $Bi_{2-x}Sb_xTe_{3-y}Se_y$ (n-BSTS and p-BSTS) after deposition of F4-TCNQ and TCNQ.

**FIGURE**

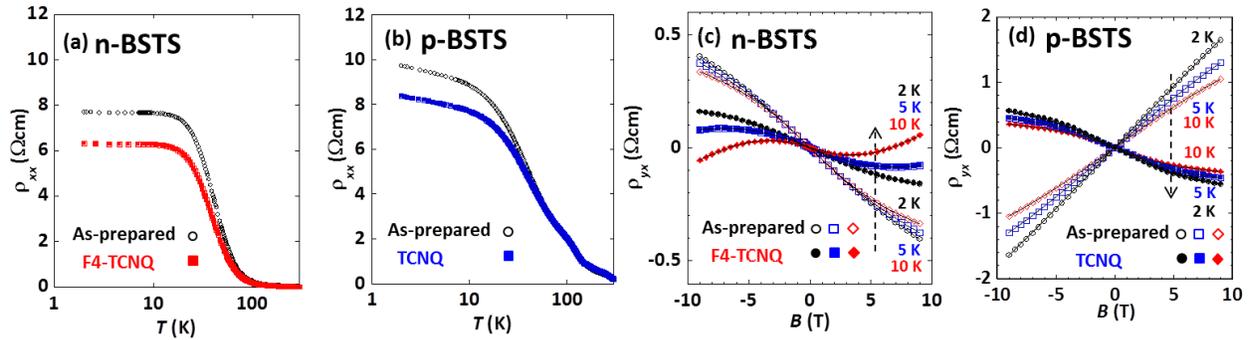

**Figure1.** Temperature dependence of electrical resistivity ($\rho_{xx}$) for (a) n-type $Bi_{2-x}Sb_xTe_{3-y}Se_y$ (n-BSTS) and (b) p-type $Bi_{2-x}Sb_xTe_{3-y}Se_y$ (p-BSTS). Magnetic field (B) dependence of Hall resistivity ($\rho_{yx}$) at temperatures between 2 and 10 K for (c) n-BSTS and (d) p-BSTS. Black, blue and red points in color correspond to data obtained at 2, 5, and 10 K, respectively. Open symbols indicate data obtained before deposition of F4-TCNQ or TCNQ and closed symbols indicate data after deposition of F4-TCNQ or TCNQ. The inset shows the results of fitting of $\rho_{yx}$. Solid lines indicate the results of fitting employing a two-carrier-type semiclassical-model (See supporting information).

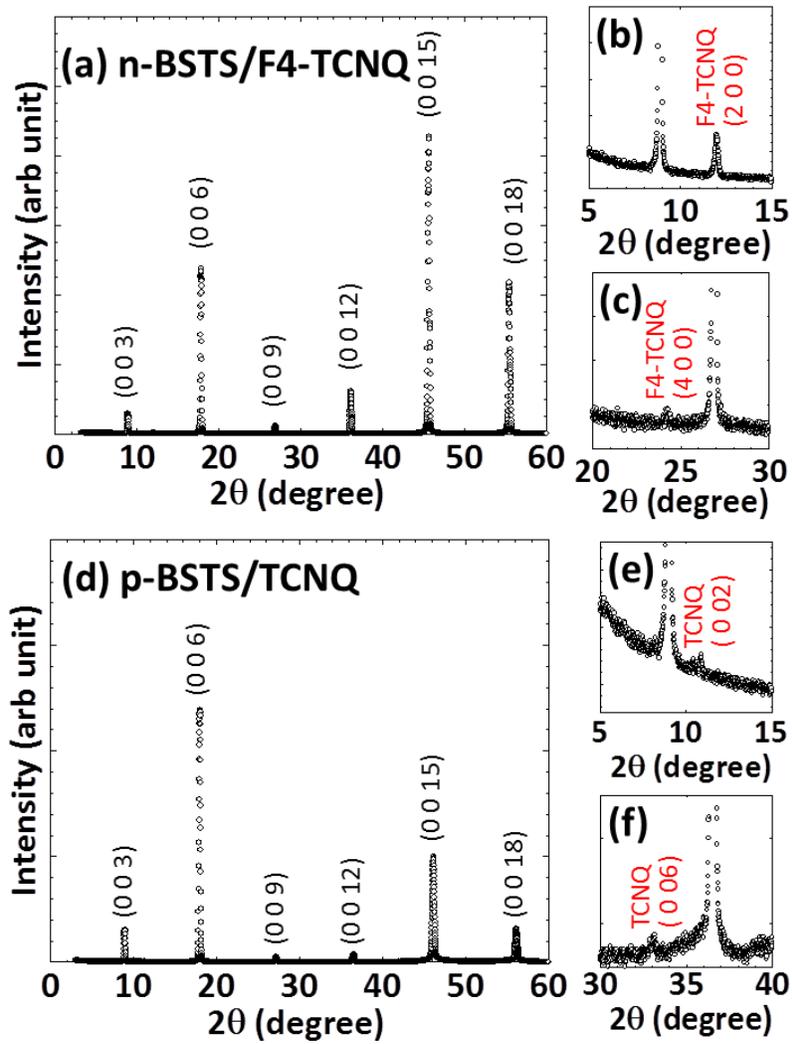

**Figure 2.** X-ray diffraction profiles for (a) n-BSTS and (d) p-BSTS along the c-axis direction after deposition of F4-TCNQ or the TCNQ. (b), (c), (e), (d) show magnified plots of (a) and (d) with visible diffraction profiles of F4-TCNQ or TCNQ on BSTS single crystals, respectively.

TABLE

**(a) n-BSTS**

| As-prepared | | | | F4-TCNQ deposition | | | |
|---|---|---|---|---|---|---|---|
| $n_1$ (1/cm$^2$) | $\mu_1$ (cm$^2$V$^{-1}$S$^{-1}$) | $n_2$ (1/cm$^2$) | $\mu_2$ (cm$^2$V$^{-1}$S$^{-1}$) | $n_1$ (1/cm$^2$) | $\mu_1$ (cm$^2$V$^{-1}$S$^{-1}$) | $n_2$ (1/cm$^2$) | $\mu_2$ (cm$^2$V$^{-1}$S$^{-1}$) |
| (n) 1.1×10$^{14}$ | 30 | (n) 8.5×10$^{10}$ | 1234 | (p) 1.0×10$^{14}$ | 41 | (n) 2.1×10$^{11}$ | 1126 |

**(b) p-BSTS**

| As-prepared | | | | TCNQ deposition | | | |
|---|---|---|---|---|---|---|---|
| $n_1$ (1/cm$^2$) | $\mu_1$ (cm$^2$V$^{-1}$S$^{-1}$) | $n_2$ (1/cm$^2$) | $\mu_2$ (cm$^2$V$^{-1}$S$^{-1}$) | $n_1$ (1/cm$^2$) | $\mu_1$ (cm$^2$V$^{-1}$S$^{-1}$) | $n_2$ (1/cm$^2$) | $\mu_2$ (cm$^2$V$^{-1}$S$^{-1}$) |
| (n) 2.0×10$^{13}$ | 60 | (p) 3.3×10$^{12}$ | 525 | (n) 3.9×10$^{14}$ | 34 | (n) 7.8×10$^{11}$ | 307 |

Table1. Carrier numbers (n) and mobilities ($\mu$) estimated from the fitting of magnetic field (B) dependence of Hall resistivity $\rho_{yx}$ at 2 K using the two-carrier type semiclassical model before and after deposition of F4-TCNQ or TCNQ for (a) n-type Bi$_{2-x}$Sb$_x$Te$_{3-y}$Se$_y$ (n-BSTS) and (b) p-type Bi$_{2-x}$Sb$_x$Te$_{3-y}$Se$_y$ (p-BSTS).

**Supporting Information**.

Hall resistivity for 2- and 3-carrier-type semiclassical model.

"This material is available free of charge via the Internet at http://pubs.acs.org."

AUTHOR INFORMATION

**Corresponding Author**

* Yoichi Tanabe

youichi@sspns.phys.tohoku.ac.jp

+81-22-217-6173

Department of Physics, Graduate School of Science, Tohoku University, Aoba, Aramaki, Aoba-ku, Sendai, 980-8578, Japan

** Khuong K. Huynh

khuong@sspns.phys.tohoku.ac.jp

+81-22-217-6173

WPI-Advanced Institutes of Materials Research, Tohoku University, 2-1-1 Katahira, Aoba-ku, Sendai, 980-8577, Japan

*** Katsumi Tanigaki

tanigaki@sspns.phys.tohoku.ac.jp

+81-22-217-6173

Department of Physics, Graduate School of Science, Tohoku University, Aoba, Aramaki, Aoba-ku, Sendai, 980-8578, Japan/ WPI-Advanced Institutes of Materials Research, Tohoku University, 2-1-1 Katahira, Aoba-ku, Sendai, 980-8577, Japan

## AUTHOR CONTRIBUTIONS

**Y. T., R. N., S. H., G. M., J. X. and S. H. have synthesized the samples. Y. T. and K. K. H. have carried out the transport measurements. Y. T., K. K. H., K. T. have analyzed and interpreted the transport data.**

## FUNDING SOURCES

The authers declare no competing financial interest.

## ACKNOWLEDGMENTS

The research was partially supported by Scientific Research on Priority Areas of New Materials Science using Regulated Nano Spaces, the Ministry of Education, Science, Sports and Culture, Grant in Aid for Science, and Technology of Japan and Grant-in-Aid for Young Scientists (B) (23740251). This work was partly supported by World Premier International Research Center Initiative (WPI), MEXT.

## REFERENCES

(1) Hasan, M. Z.; Kane, C. L. Colloquium: Topological Insulators. *Rev. Mod. Phys*. **2010**, *82*, 3045-3067.

(2) Qi, X. L.; Zhang, S. C. Topological Insulators and Superconductors. *Rev. Mod. Phys*. **2011**, *83*, 1057-1110.

(3) König, M.; Wiedmann, S.; Brüne, C.; Roth, A.; Buhmann, H.; Molenkamp, L. W.; Qi, X. L.; Zhang, S. C. Quantum Spin Hall Insulator State in HgTe Quantum Wells. *Science* **2007**, *318*, 766-770.

(4) Zhang, H.; Liu, C. X.; Qi, X. L.; Dai, X.; Fang, Z.; Zhang, S. C. Topological Insulators in $Bi_2Se_3$, $Bi_2Te_3$ and $Sb_2Te_3$ with a Single Dirac Cone on the Surface. *Nat. Phys*. **2009**, *5*, 438-442.

(5) Hsieh, D.; Qian, D.; Wray, L.; Xia, Y.; Hor, Y. S.; Cava, R.J.; Hasan, M. Z. A Topological Dirac Insulator in a Quantum Spin Hall Phase. *Nature* **2008**, *452*, 970-974.

(6) Hor, Y. S.; Richardella, A.; Roushan, Xia, Y.; Checkelsky, J. G.; Yazdani, A.; Hasan, M. Z.; Ong, N. P.; Cava, R. J. p-Type $Bi_2Se_3$ for Topological Insulator and Low-Temperature Thermoelectric Applications. *Phys. Rev. B* **2009**, *79*, 195208.

(7) Zhang, J.; Chang, C. Z.; Zhang, Z.; Wen, J.; Feng, X.; Li, K.; Liu, M.; He, K.; Wang, L.; Chen, X *et al*. Y. Band Structure Engineering in $(Bi_{1-x}Sb_x)_2Te_3$ Ternary Topological Insulators. *Nat. Commun*. **2011**, *2*, 574 (2011).

(8) Ren, Z.; Taskin, A. A.; Sasaki, S.; Segawa, K.; Ando, Y. Optimizing $Bi_{2-x}Sb_xTe_{3-y}Se_y$ Solid Solutions to Approach the Intrinsic Topological Insulator Regime. *Phys. Rev. B* **2011**, *84*, 165311.

(9) Taskin, A. A.; Ren, Z.; Sasaki, S.; Segawa, K.; Ando, Y. Observation of Dirac Holes and Electrons in a Topological Insulator. *Phys. Rev. Lett.* **2011**, *107*, 016801.

(11) Arakane, T.; Sato, T.; Souma, S.; Kosaka, K.; Nakayama, K.; Komatsu, M.; Takahashi, T.; Ren, Z.; Segawa, K.; Ando, Y. Tunable Dirac Cone in the Topological Insulator $Bi_{2-x}Sb_xTe_{3-y}Se_y$. *Nat. Commun.* **2012**, *3*, 636.

(12) Ren, Z. Taskin, A. A.; Sasaki, S.; Segawa, K.; Ando, Y. Fermi Level Tuning and a Large Activation Gap Achieved in the Topological Insulator $Bi_2Te_2Se$ by Sn doping. *Phys. Rev. B* **2012**, *85*, 155301.

(13) Sacepe, B.; Oostinga, J. B.; Li, J.; Ubaldini, A.; Couto, N. J. G.; Giannini, E.; Morpurgo, A. F. Gate-Tuned Normal and Superconducting Transport at the Surface of a Topological Insulator. *Nat. Commun.* **2011**, *2*, 575.

(14) Steinberg, H.; Gardner, D. R.; Lee, Y. S.; Jarillo-Herrero, P. Surface State Transport and Ambipolar Electric Field Effect in $Bi_2Se_3$ Nanodevices. *Nano Lett.* **2010**, *10*, 5032-5036.

(15) Chen, J.; Qin, H. J.; Yang, F.; Liu, J.; Guan, T.; Qu, F. M.; Zhang, G. H.; Shi, J. R.; Xie, X. C.; Yang, C. L. *et al.* Gate-Voltage Control of Chemical Potential and Weak Antilocalization in $Bi_2Se_3$. *Phys. Rev. Lett.* **2010**, *105*, 176602.

(16) Checkelsky, J. G.; Hor, Y. S.; Cava, R. J.; Ong, N. P. Bulk Band Gap and Surface State Conduction Observed in Voltage-Tuned Crystals of the Topological Insulator $Bi_2Se_3$. *Phys. Rev. Lett*. **2010**, *106*, 196801.

(17) Kong, D.; Chen,Y.; Cha, J. J.; Zhang, Q.; Analytis, J. G.; Lai, K.; Liu, Z.; Hong, S. S.; Koski, K. J.; Mo, S. K. *et al*. Ambipolar Field Effect in the Ternary Topological Insulator (Bi$_x$Sb$_{1-x}$)$_2$Te$_3$ by Composition Tuning. *Nat. Nanotech.* **2011**, *6*, 705-709.

(18) Chen, Y. L.; Chu, J. H.; Analytis, J. G.; Liu, Z. K.; Igarashi, K.; Kuo, H. H.; Qi, X. L.; Mo, S. K.; Moore, R. G.; Lu, D. H. *et al*. X. Massive Dirac Fermion on the Surface of a Magnetically Doped Topological Insulator. *Science* **2010**, 329, 659-662.

(19) Kong, D.; Cha, J. J.; Lai, K.; Peng, H.; Analytis, J. G.; Meister, S.; Chen, Y.; Zhang, H. J.; Fisher, I. R.; Shen, Z. X. *et al*. Rapid Surface Oxidation as a Source of Surface Degradation Factor for Bi$_2$Se$_3$. *ACS Nano* **2011**, *5*, 4698-4703.

(20) Yuan, H.; Liu, H.; Shimotani, H.; Guo, H.; Chen, M.; Xue, Q.; Iwasa, Y. Liquid-Gated Ambipolar Transport in Ultrathin Films of a Topological Insulator Bi$_2$Te$_3$. *Nano Lett.* **2011**, *11*, 2601-2605.

(21) Kim, D.; Cho, S.; Butch, N. P.; Syers, P.; Kirshenbaum, K.; Adam, S.; Paglione, J.; Fuhrer, M. S. Surface Conduction of Topological Dirac Electrons in Bulk Insulating Bi$_2$Se$_3$. *Nat. Phys.* **2012**, 8, 459-463.

(22) Wang, J.; Chen, X.; Zhu, B. F.; Zhang, S. C. Topological p-n Junction. *Phys. Rev. B* **2012**, *85*, 235131.

(23) Katsnelson, M. I.; Novoselov, K. S.; Geim, A. K. Chiral Tunnelling and the Klein Paradox in Graphene. *Nat. Phys.* **2006**, *2*, 620-625.

(24) Cheianov, V. V.; Fal'ko, V.; Altshuler, B. L. The Focusing of Electron Flow and a Veselago Lens in Graphene p-n Junctions . *Science* **2007**, *315*, 1252-1255.

(25) Aristizabal, C. O.; Fuhrer, M. S.; Butch, N. P.; Paglione, J.; Appelbaum, I.; Towards Spin Injection from Silicon into Topological Insulators: Schottky Barrier between Si and $Bi_2Se_3$. *Appl. Phys. Lett.* **2012**, 101, 023102.

(26) Nagao, J.; Hatta, E.; Mukasa, K.; *Evaluation of Metal-$Bi_2Te_3$ Contacts by Electron Tunneling Spectroscopy.* Proceedings of the XV International Conference on Thermoelectrics. Pasadena, California, USA, **1996**.

(27) Gao, W. Y.; Kahn, A. Controlled Doping of the Hole-Transport Molecular Material with Tetrafluorotetracyanoquinodimethane. *J. Appl. Phys.* **2003**, *94***,** 359-366.

(28) Milián, B.; Amérigo, R. P.; Viruela, R.; Orti, E. On the Electron Affinity of TCNQ. *Chem. Phys. Lett.* **2004**, *391*, 148-151.

(29) Emge, T. J.; Maxfield. M.; Cowan, D. O.; Kistenmacher, T. J. Solution and Solid State Studies of Tetrafluoro-7,7,8,8-Tetracyano-p-Quinodimethane, TCNQF4. Evidence for Long-Range Amphoteric Intermolecular Interactions and Low-Dimensionality in the Solid State Structure. *Mol. Cryst. Kiq. Cryst.* **1981**, *65*, 161-178.

(30) Long, R. E.; Sparks, R. A.; Trueblood, K. N. The Crystal and Molecular Structure of 7,7,8,8-Tetracyanoquinodimethane. *Acta. Cryst.* **1965**, *18*, 932-939.

(31) Chang, C. Z.; He, K.; Wang, L. L.; Ma, X. C.; Liu, M. H.; Zhang, Z. C.; Chen, X.; Xue, Q. K. Growth of Quantum Well Films of Topological Insulator $Bi_2Se_3$ on Insulating Substrate. *Spin* **2011**, *1*, 21-25.

(32) Wang, G.; Zhu, X.; G. Sun, Y. Y.; Li, Y. Y.; Zhang, T.; Wen, J.; Chen, X.; He, K.; Wang, L. L.; Ma, X. C. *et al.* Topological Insulator Thin Films of $Bi_2Te_3$ with Controlled Electronic Structure. *Adv. Mater.* **2011**, *2*, 2929-2932.

(33) Zhang, Y.; He, K.; Chang, C. Z.; Song, C. L.; Wang, L. L.; Chen, X.; Jia, J. F.; Fang, Z.; Dai, X.; Shan, W. Y. *et al.* Crossover of the Three-Dimensional Topological Insulator $Bi_2Se_3$ to the Two-Dimensional Limit. *Nat. Phys.* **2010**, *6*, 584-588.

(34) Bianchi, M.; Guan, D.; Bao, S.; Mi, J.; Iversen, B.; B. King, D. C. P.; Hofmann, P. Coexistence of the Topological State and a Two-Dimensional Electron Gas on the Surface of $Bi_2Se_3$. *Nat. Commun.* **2010**, *1*, 128.

(35) Qi, X. L.; Hughes, T. L.; Zhang, S. C. Topological Field Theory of Time-Reversal Invariant Insulators. *Phys. Rev. B* **2008**, *78*, 195424.

(36) Nomura, K.; Nagaosa, N. Surface-Quantized Anomalous Hall Current and the Magnetoelectric Effect in Magnetically Disordered Topological Insulators. *Phys. Rev. Lett.* **2011**, *106*, 166802.